\documentclass[preprint]{./acm_proc_article-sp}
\usepackage{epsfig}
\usepackage[square,comma,numbers,sort&compress]{natbib}
\usepackage{multirow}

\newcommand{\UPLB}{University of the Philippines Los Ba\~{n}os}
\newcommand{\SN}{\mathcal{SN}}
\newcommand{\adjmat}{\mathcal{M}}
\newcommand{\speed}{\mathcal{V}}
\newcommand{\width}{\mathcal{W}}
\newcommand{\degree}{\Delta}
\newcommand{\power}{\varphi}
\newcommand{\info}{\mathcal{I}}
\newcommand{\digg}{{\tt digg.com}\ }

\begin{document}
\title{Information Spread Over an\\Internet-mediated Social Network:\\Phases, Speed, Width, and Effects of Promotion}
\numberofauthors{1}
\author{
\alignauthor Abigail C. Salvania and Jaderick P. Pabico\titlenote{Author contributions: A.C.S. implemented the computational solution, conducted the computational experiments, and prepared and edited the final paper. J.P.P. formulated the computational solution, conducted the statistical analysis, and wrote the final paper.}\\
   \affaddr{Institute of Computer Science}\\
   \affaddr{\UPLB}\\
   \affaddr{College 4031, Laguna}
}
\date{}
\maketitle

\begin{abstract}

In this study, we looked at the effect of promotion in the speed and width of spread of information on the Internet by tracking the diffusion of news articles over a social network. Speed of spread means the number of readers that the news has reached in a given time, while width of spread means how far the story has travelled from the news originator within the social network. After analyzing six stories in a 30-hour time span, we found out that the lifetime of a story's popularity among the members of the social network has three phases: Expansion, Front-page, and Saturation. Expansion phase starts when a story is published and the article spreads from a source node to nodes within a connected component of the social network. Front-page phase happens when a news aggregator promotes the story in its front page resulting to the story's faster rate of spread among the connected nodes while at the same time spreading the article to nodes outside the original connected component of the social network. Saturation phase is when the story ages and its rate of spread within the social network slows down, suggesting popularity saturation among the nodes. Within these three phases, we observed minimal changes on the width of information spread as suggested by relatively low increase of the width of the spread's diameter within the social network. We see that this paper provides the various stakeholders a first-hand empirical data for modeling, designing, and improving the current web-based services, specifically the IT  educators for designing and improving academic curricula, and for improving the current web-enabled deployment of knowledge and online evaluation of skills.
\end{abstract}


\section{Introduction} 

News articles are an important form of social communications because they provide records of recent history that affected the community, as well as forecasts of future events (e.g., weather and business forecasts) that could potentially affect the members. Even before the information age, and more so during the information age, news articles have become an integral part of human's daily lives that more and more people have become dependent on these type of information for proper and timely decision-making processes. Since news have become an indispensible tool for humans, their spreading across the social network also plays a significant role in a variety of human affairs. For example, they can shape the public opinion in a country~\citep{galam03}, greatly impact financial markets~\citep{kosfeld05}, cause citizens to take actions on some issues~\citep{carr10}, effect the spread of technological innovations~\citep{rogers95,strang98}, and speed up word-of-mouth effects in marketing~\citep{domingos01,kempe03,leskovec06}. The information content of news articles can range from simple reporting of an event containing basic information, to advanced propaganda and marketing of various stakeholders, such as that of private citizens who selectively edit news items and then broadcasts them via their own news conferences to further their cause, or the goverment which filters information for national security reasons. Mechanisms for spreading information form the basis of the Internet which have become, in recent years, the playground for testing and evaluating an important class of digital communication protocols. Measuring the speed and width of information spread over an Internet-mediated social network, as well as quantifying the effect of promotion on these measures, may provide various scientists and engineers a first-hand empirical result to aid them in modeling, designing, and improving the current Internet services. Likewise, this will aid IT technical trainors and academic educators in designing  innovative curricula, as well as use and improve tools for web-enabled deployment and evaluation of knowledge and skills.

Researchers in the study of information spread have assumed that information $\info$ diffuses over a social network~$\SN$ from a person $v_0\in\SN$ to another person $v_j\in\SN$ following an epidemic-like dynamics~\citep{dodds04}, and spreading to other person-members of~$\SN$ in a short number of steps according to the {\em small-world} principles~\citep{travers69,watts98}. Despite the numerous studies in the past years under this scientific domain, the process of documenting, and thereby tracing, the dissemination of a single piece of news article among the members of $\SN$, particularly for nationally and globally-sized $\SN$s, has been rather difficult, not only due to the sheer size of the study subject, but moreso due to the inherrent dynamism of the subject~$\SN$. Thus, researchers have not yet agreed whether the spread of $\info$ over a very large and dynamic~$\SN $ really follows a rapid, epidemic-like fanning out, or it follows another process that could be simpler or more complex than that of the epidemics model. 

In this effort, to characterize the spread of $\info$ over an Internet-mediated $\SN$, we followed the spread of~$\info$ through a social web over its perceived lifetime while measuring its speed~$\speed$ and width~$\width$ of spread, as well as the effect of the  promotion of $\info$ on these metrics. We found out that the lifetime of  the popularity of~$\info$ among the members of~$\SN$ undergoes three phases, namely, expansion, front-page, and saturation. We measured the respective $\speed$ and $\width$ of each phase and found out the following:
\begin{enumerate}
\item {\bf Expansion Phase} is when an $\info$ is published by a member $v_0\in\SN$ and it spreads to other members $v_i\in\SN$, $\forall i>0$, who are directly related in one way to $v_0$. The speed of spread ranges from 1\% to 12\% of the total size of $\SN$ per hour.
\item {\bf Front-page Phase} is when a news aggregator promotes $\info$ resulting in faster $\speed$ among the connected members of $\SN$, and at the same time spreading $\info$ to members outside of this connected members, which is rather contrary to previously developed models~\citep{LibenNowell}. During this time, $\speed$ ranges from 3\% to 23\% of the total size of $\SN$ per hour.\item {\bf Saturation Phase} happens when $\info$ ages and $\speed$ slows down suggesting saturation among the members of $\SN$. This time, $\speed$ ranges from an hourly 1\% to 4\% of the total size of $\SN$. 
\end{enumerate}

\section{Conceptual Framework}

The social network~$\SN$ is abstractly represented as a graph $G(V,E)$ composed of social members, such as humans, represented as a set~$V$ of $n$~vertices $v_0, v_1, \dots, v_{n-1}$. The relationship of a member~$v_i$ to any other member~$v_j$, $\forall i\not=j$, is represented as an edge $(i, j)$. The relationship of all members with the other members is a set~$E$ of edges, and can be mathematically represented as an $n\times n$ adjacency matrix~$\adjmat$. The matrix element $\adjmat_{i,j}=1$ if~$v_i$ is related in someway to~$v_j$. Otherwise, $\adjmat_{i,j}=0$. We note here that the relationship is symmetric such that~$v_i$ having a relationship with~$v_j$ also means~$v_j$ having a relationship with~$v_i$. Thus, $\adjmat_{i,j} = \adjmat_{j,i}$. Without loss of generality, we set $\adjmat_{i,i}=0, \forall i$. For relationships whose degrees~$r$ can be quantified, the edge $(i,j)$ is labeled by~$r$, while $0\le\adjmat_{i,j}\le 1$, where $\adjmat_{i,j}=r>0$ if~$v_i$ is related to~$v_j$ to some degree, otherwise $\adjmat_{i,j}=0$. In this case, the relationship is no longer symmetric. Under these degreed relationships, however, measuring~$r$ is quite difficult to achieve objectively for most real world~$\SN$s. In this paper, we assumed that $\adjmat_{i,j}$ is binary, while we will consider the more general $0\le\adjmat_{i,j}\le 1$ in the future.

The degree $\degree_i$ of a vertex $v_i$ counts the number of humans that member $v_i$ has a relationship with. Thus, $\degree_i=\sum_{j=1}^n \adjmat_{i,j}$. Recently, the frequency distribution $\rho(\degree)$ of the degree in $\SN$ has been found by various researchers~\citep{barabasi99,albert02,barabasi03} to asymptotically follow the power law distribution of the form $\rho(\degree)=\alpha\times\degree^\power$. For social networks, and all other biological networks, the power usually takes the value $-3\le\power\le -2$. Having $\rho(\degree) \sim \alpha\times\degree^\power$ makes $\SN$ scale-free~\cite{albert02}.

An information~$\info$, usually in the form of a news article, spreads through~$\SN$ in the following manner. Let $S(v_i)$ be a function that represents the state~$S$ of a vertex $v_i\in V$. The state $S(v_i)=0$ if~$v_i$ has no knowledge yet of the presence of~$\info$. This, intuitively, means that~$\info$ has not reached~$v_i$ yet. However, $S(v_i)=1$ if~$\info$ has reached~$v_i$. Note here that~$S$ only changes from~0 to~1 and not the other way around. At time $t_0$, $\info$ originates from a source member $v_0\in\SN$, at which time $S(v_0)=1$ while all the others have $S(v_k)=0, \forall k>0$. During the succeeding time periods, $t_1, t_2, \dots, t_k$, $k>0$, $\info$ spreads through $\SN$ via some of the  edges that are connected to $v_0$. In turn, $\info$ spreads through members $v_i$, for some $i>0$, and $\adjmat_{i,0}=1$, during which $S(v_i)=1$. The sum $\sum_i S(v_i)$ within a time interval $\delta t$ is the rate of spread $\speed$ of $\info$ in $\SN$.

For the connected components of $\SN$, there exists a vertex $v_i\in\SN$ whose  state $S(v_i)=1$. From $v_0$, we can find a shortest path $D_{0,i}$ out of potentially many paths from $v_0$ to $v_i$ whose intermediate vertices $v_j$, $0,i\not=j$, have $S(v_j)=1$. The path distance can be computed as $D_{0,i}=\sum S(v_j)$. Note that $D_{0,i}=D_{i,0}$. The longest among all the shortest paths from $v_0$ to $v_i$ is the radius of spread of $\info$ through the connected components of $\SN$. Using the all-pairs shortest paths algorithms by~\citet{floyd62} and by~\citet{johnson77}, we can approximate the~$\width$ by computing for the longest shortest path of the connected component within~$\SN$, but considering only those paths that will traverse~$v_0$. 

With this scenario, we may quantify the spread of $\info$ over $\SN$ in two ways. First is the speed $\speed$ at which the information has reached the members of the community over a time period  (Equation~\ref{eqn:speed}), and second is the diameter width~$\width$ of the component of~$\SN$ at which the information has spread from the source~$v_0$. 

\begin{equation}
\speed = \frac{\sum_{i=0}^{n-1} S(v_i)}{\delta t}\label{eqn:speed}
\end{equation}

\subsection{Internet-mediated social networks}

In its early history, the Internet primarily served a worldwide research community while its developers worked on improving communication protocols and inventing new ones to support emerging services. Even though the growth of Internet eased as a result of the bursting of the dot-com bubble in the early 2000's, it still has become an integral part of everyday life for ordinary people~\cite{Brownlee} due to many reasons brought about by the effect of technology in people's lives. One among the many reasons is that it has become an obiquitous tool to meeting, communicating and collaborating with other people resulting in the formation of the {\em Social Web}.

\begin{figure}[hbt]
\centering\epsfig{file=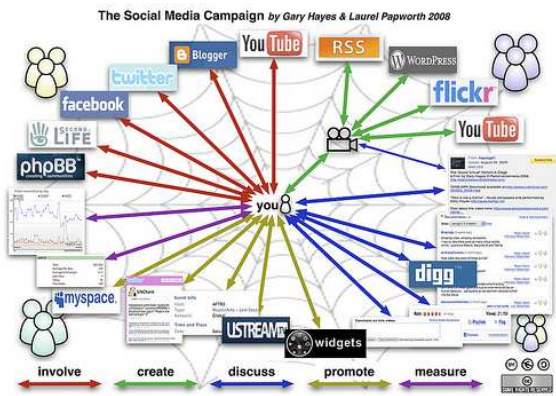, width=3in}
\caption{A popular visualization of the conceptual framework of the Social Web by~\citet{hayes08} out of the many visualizations that exist.}\label{fig:social-web}
\end{figure}

The {\em Social Web} is a label that includes both the social networking sites such as MySpace and Facebook, and the social media sites such as Digg and Flickr.  Figure~\ref{fig:social-web} shows one of the many visualizations of the conceptual framework of the social web. The social web, in its many forms, has been tagged as one of the many forces in changing the way content is created and distributed in the Internet. This change was further expidiated through the proliferation of web-based authoring tools that enable users to rapidly publish content, such as stories, opinion pieces on weblogs, photographs and videos on Flickr and YouTube, advices on Yahoo Answers, and web discoveries on Del.icio.us and Furl. In effect, user-generated content has fueled the rapid expansion of the Web, accounting for much of the new Web content in the Internet. In addition to allowing users to share content, social web sites also include a social networking component which allows users to mark other users as friends or contacts. They also provide an interface to track their friends' activities~\citep{Brownlee}. These components have a significant impact on the Internet traffic distribution. Understanding the feature and characteristics of these sites is crucial to network traffic engineering and to sustain the development of this new generation of service~\citep{Arevalo}. 

\subsection{Recent studies in information spread}

Different studies have been conducted in the past to trace how $\info$ spreads on the Internet. In one of their recent studies, \citet{LibenNowell} found out that the path followed by $\info$ was narrow and is characterized by a very deep tree-like pattern rather than fanning out like the small world principle suggests. They observed this particular behavior after reconstructing the paths followed by $\info$ in a global-scale Internet chain letter data. They reconstructed the paths by tracing the signatories of the chain letter data. In similar studies conducted by~\citet{Lerman} and~\citet{Xu}, where they independently observed the spread of $\info$ on different social news aggregators, they found out that~$\SN$ play a significant role in the promotion of $\info$ among the members of~$\SN$. In fact, \citet{Xu} even found that the networks of related $\info$ are characterized by a short path length linking two similar $\info$s together. This suggests that the inferred~$\SN$ in social news aggregators follow the small-world principle.

\subsection{Case study: \digg}

To characterize the spread of $\info$ over an Internet-mediated $\SN$, we followed the spread of~$\info$ through a social web called \digg over a period of time. We were able to measure the respective $\speed$ and $\width$ of spread of six stories $\{\info_1, \info_2, \dots, \info_6\}$, as well as the effect of the  promotion of the stories on $\speed$ and $\width$ of these. The social website \digg is a digital media democracy that allows its members to submit $\info$ in the form of news stories that they discovered from the Internet. When a member $v_0\in\SN$ submits a news article, \digg will list $\info$ under the submitter's name, as well as put $\info$ to a repository web page called ``Upcoming Stories,'' where other members $v_i\in\SN$, $\forall i>0$, can find $\info$ and, if they like it, add a ``digg'' to it. Adding a ``digg'' to an~$\info$ is synonymous to ``voting'' for it. When a member $v_i$ diggs an $\info$, that $\info$ is saved to $v_i$'s history. At the same time, a so-called \digg number is shown next to each $\info$'s title. This number simply counts how many members have digged $\info$ in the past. Most diggs come from the network neighbors of~$v_0$. New diggs might also attract additional diggs from the neighbors of the diggers, and so on. 

\begin{figure}[hbt]
\centering\epsfig{file=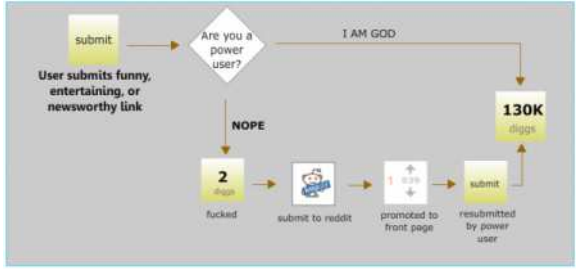, width=3in}
\caption{Users have speculated how \digg promotes~$\info$ into its front page because the exact algorithm has been kept secret in an effort to protect it from systematic cracking by malware software writers. This figure by~\citet{finch09} is one of those speculations.}\label{fig:digg-promotion-flow-chart}
\end{figure}

If a submission fails to receive enough diggs within a certain time period, it eventually falls out of the ``Upcoming'' section. However, if it earns a critical mass of diggs in a short span of time, it becomes popular and is promoted to the \digg's front page. Most members read only the front page stories, thus, getting on the front page greatly increases a specific $\info$'s visibility. Promoted $\info$ is selected by \digg's algorithms to be displayed on the front page, but the exact formula for how a story is selected is kept secret and changes periodically (Figure~\ref{fig:digg-promotion-flow-chart}). This prevents users from ``gaming the system,'' a systematic crack by black hat hackers for the sole purpose of promoting advertisement or spam. However, it appears that the algorithm takes into account, among other things, the number of votes received by $\info$~\cite{LermanK}.

\section{Methodology}

Instead of obtaining the data from the social web operator, we crawled our target~$\SN$ by accessing the public web interface provided by the \digg API. Our crawler was developed using Perl scripts and utilizes the Linux command line programs grep and wget. The public interface provided by the \digg API outputs an array of $p$~pages containing $n$~accounts. Each of the first $p-1$~pages lists $100$ unique accounts while the last page lists $(n \mod 100)$ accounts. The default account per page is 10 but the maximum allowed is 100, which can be set by assigning the proper variable in the URL. We automatically crawled the $p$~pages by changing the proper parameters in each URL, a method that we adopted from the work done by~\citet{Arevalo}.

Our crawler first extracts the diggers or voters of a given $\info$. For each $v_i\in\SN$, our crawler then extracts the list of $v_i$'s friends. To be able to view the diggers of a story, the \digg API provided the URL that shows the list of diggers of a certain story, including the digg date, story id, digger's user id and name, and the status of the story~\cite{DapiA}. To be able to view the list of friends of the diggers of a story, \digg API also provided the URL which contains the friends' user name, number of profile views, and registration date of user~\cite{DapiB}. We stored these information in separate database tables named ``digger'' and ``friends,'' respectively. The ``digger'' table includes the user name, user id, story id, and digg date as attributes. At the end of the crawl, for each $\info$, the ``digger'' table will contain $n$~unique records corresponding to each digger discovered by the crawler. The ``friends'' table takes the user name of the diggers as well as the user name of his friends. The digger's user name in the ``digger'' table is used as a foreign key for the other table containing his friends. The crawler disregards inactive users, which is already flagged by the \digg API. The trace for each story is stored in a different database.  Information about the $\info$, like its title, story id, name of submitter, date the story was submitted, and date the story was promoted were also taken into account and stored. All of these information can be viewed on the URL provided by the \digg API~\cite{DapiC}.

In representing the~$\SN$ of \digg users, we treated the diggers as vertices while the relationships between them as edges. We considered the first vertex $v_0\in\SN$ as the submitter of~$\info$, while the succeeding other vertices $v_i$, $\forall i>0$, were the diggers of~$\info$. Our crawler checks for the  succeeding digger's user name if it has a relationship with the previous diggers. If the succeeding digger $v_k\in\SN$ is a friend of the previous diggers $v_i$, $\forall i<k$, we connect $v_k$ to those previous diggers. We also considered the time of digging to accurately construct the evolution graph of diggers. 

We crawled the \digg $\SN$ every hour to get a snapshot of the structure of the~$\SN$. For each reading, we measured the number of diggers and network diameter of the inferred~$\SN$. We considered the speed~$\speed$ of the information spread as the number of diggers~$n_t$ at a certain time~$t$, while the width~$\width$ of information spread is the network diameter of the connected component of the graph, modified according to our needs. This means that we only considered those nodes whose $S(v_i)=1$ and those paths which will travese~$v_0$. 

\section{Results and Discussion}

\subsection{Stories traced}

\begin{figure}[hbt]
\centering\epsfig{file=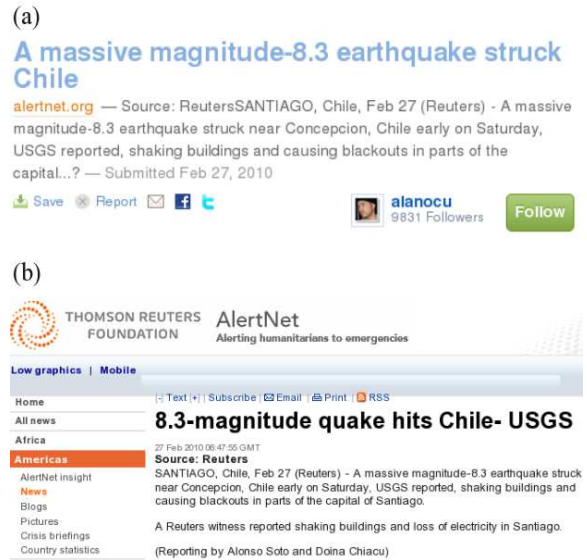,width=3in}
\caption{The front page of story $\info_3$ (a)~as seen in \digg on 27 February 2010,  and (b)~as seen in its original news website Thomson Reuters Foundation.}\label{fig:story3}
\end{figure}

We set our crawler to trace the information spread of six stories between January and February 2010. Specific information about the stories, its diggers and the list of their friends were saved in a database. These stories are listed in Table~\ref{tab:stories}, together with some statistics that we gathered after the trace of our crawler. Story $\info_3$ (Figure~\ref{fig:story3}) had the most number of diggers and the largest network diameter, while Story $\info_2$ had the least number of diggers and Story $\info_1$ had the smallest network diameter. Story $\info_3$ had the fastest rate and widest width of information spread after 30 hours. Story $\info_2$ had the slowest rate of information spread after 30 hours and Story $\info_1$ had the least width of information spread after 30 hours.

\begin{table*}[hbt]
\caption{The six stories that we traced in this study: Story title, date submitted, name of the submitter ($v_0$), size $n$ of~$\SN$ and the $\width$ after 30 hours.}\label{tab:stories}
\centering\begin{tabular}{llllrr}
\hline\hline
Story & Title & Date      & $v_0$ & $n$ & $\width$\\
      &       & Submitted &  &  & \\\hline
$\info_1$& Haterade Power Rankings Super Bowl Edition& 10 Feb 2010& racheljtm & 181 & 4 \\
$\info_2$& NSAC Manny Pacquiao's Drug Test results are Negative&16 Jan 2010 & lovelezz & 147 & 7 \\
$\info_3$& A massive magnitude 8.3 earthquake struck Chile&27 Feb 2010 & alanocu & 1,378 & 10 \\
$\info_4$& China warns Obama not to meet Dalai Lama 2&02 Feb 2010 & divinediva & 989 & 9 \\
$\info_5$& Please Help The People of Haiti&16 Jan 2010 & jayadelson & 661 & 7 \\
$\info_6$& Ohno's Wins 7th Medal and Sets US Record&21 Feb 2010& robinbal & 177 & 5 \\
\hline\hline
\end{tabular}
\end{table*}

\subsection{Phases in the life of a news article}

\begin{figure*}[hbt]
\centering\epsfig{file=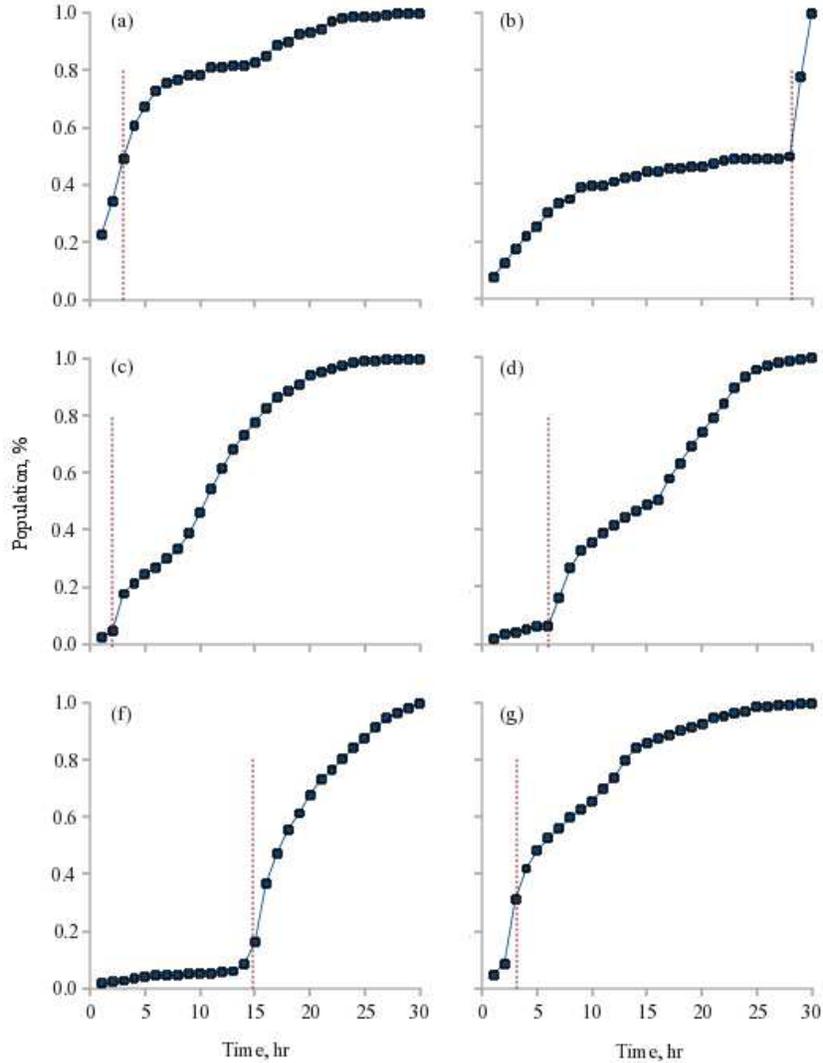,width=4.35in}
\caption{Percentage of population reached by various stories within the 30-hour trace: (a)~$\info_1$; (a)~$\info_2$; (b)~$\info_3$; (c)~$\info_4$; (d)~$\info_5$; and (e)~$\info_6$. Each red vertical line marks the time when the story was promoted to the front-page.}\label{fig:phases}
\end{figure*}

Figure~\ref{fig:phases} shows the respective traces of $\info_i$, $\forall i=1, \dots, 6$ in terms of the percentage of the population of~$\SN$. From the figures, we see three prominent patterns: The behavior of the plot before the promotion, after promotion, and an ageing pattern. We termed these phases as follows:
\begin{enumerate}
\item Expansion Phase: This phase starts when $\info$ is published and the article spreads from $v_0$ to other members of $\SN$ who are at least separated from $v_0$ by at least $\degree=3$.
\item Front-Page Phase: This phase happens for some $\info$ wherein a news aggregator at time~$t$ promotes~$\info$ into its front page. This gives boost to the spread of the article outside of the connected component of $v_0$ at~$t$ resulting to faster speed of spread.
\item Saturation Phase: This phase is considered when the $\info$ ages and its popularity among the members of the $\SN$ died down, that is $\delta S/\delta t<0.5$. From the figure, this usually happens when about 90\% of the population have already read about~$\info$.
\end{enumerate}

Stories $\info_1$ and $\info_6$ seemed to follow the same pattern, which is characterized by an upward slope $\delta S/\delta t>0.5$ during the early hours of its life, and then leveling off with $\delta S/\delta t<0.5$. It seemed that these two stories have two phases because there is not much break on the spead of spread of $\info$ before and after the promotion. However, we note here that these two news articles are both sports-related,  both talk about the sports in the US where most of the \digg members reside, and that the news articles were published when the respective sports are being talked about by the mainstream media. This means that even though without the benefit of the front-page promotion, we expected that these two news items will have $\delta S/\delta t>0.5$ because the respective topics have already been promoted elsewhere. 

Stories $\info_2$ through $\info_5$ show a more pronounced break between the phases before and after promotion. For example, $\info_2$ which been promoted about 28 hours after its publication seemed to follow a natural ageing phase after about the 9th hour of its life (Figure~\ref{fig:phases}b). Notice, however, that there was a considerable ``jump'' after it has been promoted to the front-page. Likewise, $\info_3$ through $\info_5$ exhibit the same jump, after their promotions on the 3rd, 7th, and 15th hours, respectively.

\subsection{Speed, width, and effects of promotion}

\begin{figure*}[hbt]
\centering\epsfig{file=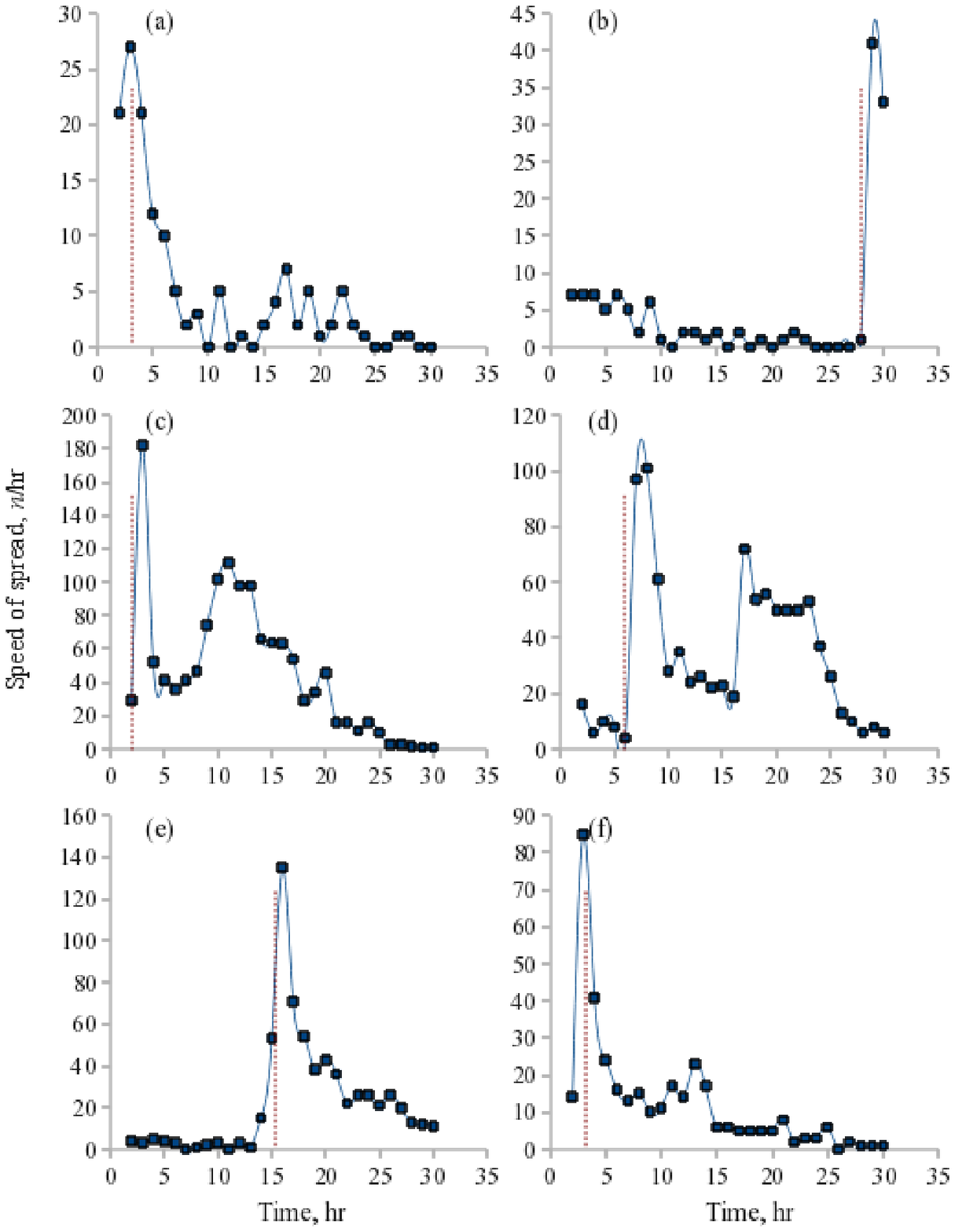,width=4.35in}
\caption{Various plots of instantaneous speed $\speed$ of spread of the stories during the 30-hour trace:  (a)~$\info_1$; (a)~$\info_2$; (b)~$\info_3$; (c)~$\info_4$; (d)~$\info_5$; and (e)~$\info_6$. Each red vertical line marks the time when the story was promoted to the front-page.}\label{fig:speed}
\end{figure*}

Figure~\ref{fig:speed} shows the plot of instantaneous speed $\speed$ of spread of each story during the 30-hour trace. Except for~$\info_1$ and~$\info_6$, all other stories exhibit a tremendous rise in~$\speed$ after their respective promotions to \digg front-page. Stories~$\info_1$ and~$\info_6$ did not show this pattern due to the same characteristics enumerated above. Before the promotion, the spread of a story loses speed as clearly shown in the respective behaviours of $\info_2$, $\info_4$, and $\info_5$. A single peak lasting about one hour is observed just after promotion. Some others have multiple peaks, such as~$\info_3$ and~$\info_4$, but the height of the peaks is no greter than the peak just after promotion. The $\speed$ suffered a downward patterns right after every peak, following the downward path just before the second peak (see for example that of $\info_4$ which has a more pronounced pattern than the others), suggesting saturation of the population.

\begin{figure*}[hbt]
\centering\epsfig{file=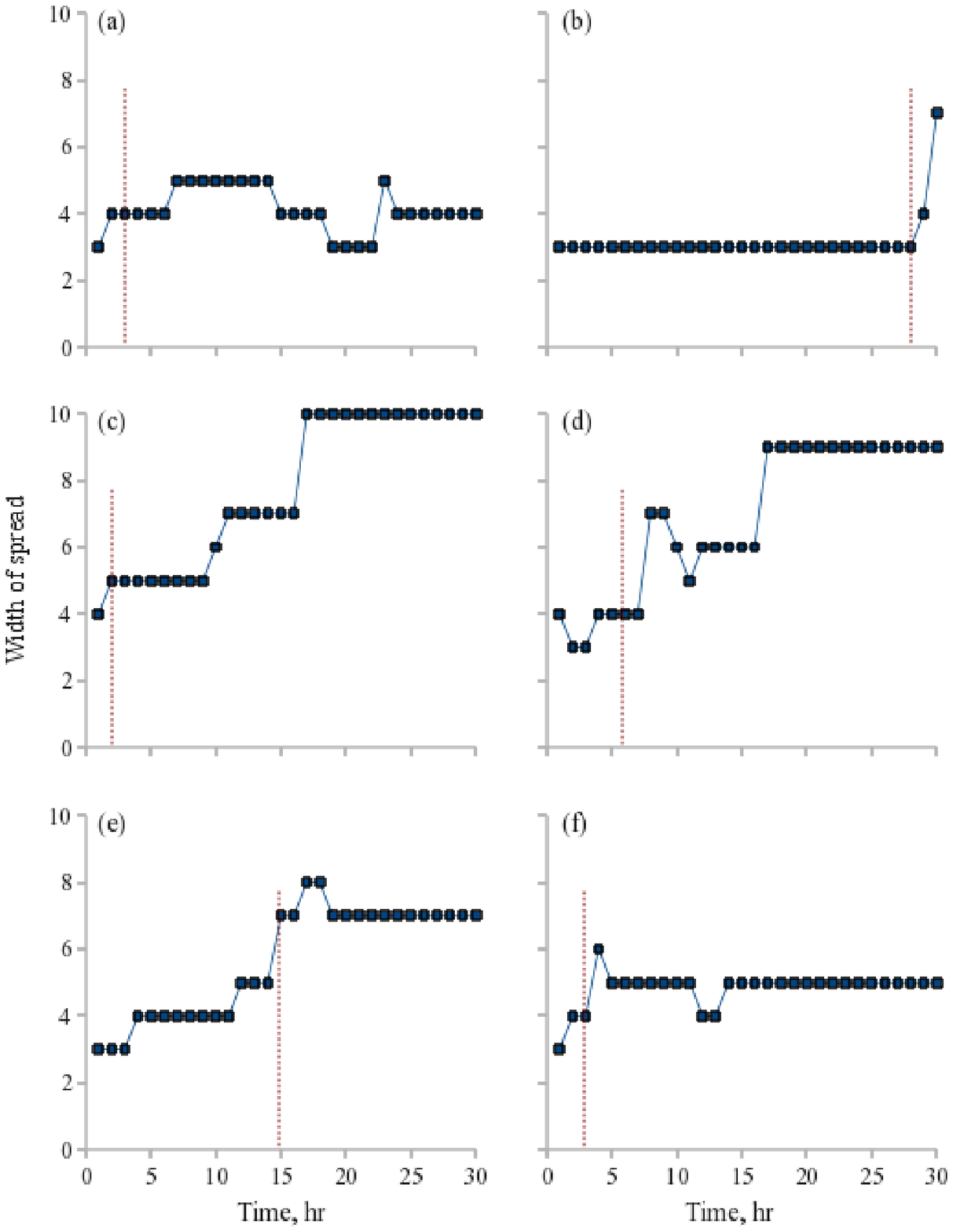,width=4.35in}
\caption{Various plots of width $\width$ of spread of the stories during the 30-hour trace:  (a)~$\info_1$; (a)~$\info_2$; (b)~$\info_3$; (c)~$\info_4$; (d)~$\info_5$; and (e)~$\info_6$. Each red vertical line marks the time when the story was promoted to the front-page.}\label{fig:width}
\end{figure*}

Figure~\ref{fig:width} shows the various plots of the evolution of the width~$\width$ of spread of each story during the 30-hour trace. The most notable value among the plots is that the minimum $\width=3$. The physical implication of this minimum $\width$ value is that at this width, $\info$ has only reached the immediate neighbors of $v_0$, wherein at least one neighbor $v_i$ is not directly connected to the other neighbors of $v_0$. One notable story is~$\info_2$, wherein it spread only to the immediate neighbors of the submitter $v_0$ during the first 28 hours of the story's life. However, after the story's promotion to the front-page, the width of spread expanded up to seven members within a two-hour span. Only~$\info_2$ and~$\info_6$ experienced an immediate expansion of the width of spread after promotion, width $info_2$ expanding by one member and $\info_6$ by two. All the other stories continued on with their respective widths until at least one hour after the front-page promotion. The physical implication of a decreasing $\width$ is that $\info$ has reached a member that caused the longest shortest path to shorten. 

\subsection{Evolution of information spread}

\begin{figure*}[htb]
\centering\epsfig{file=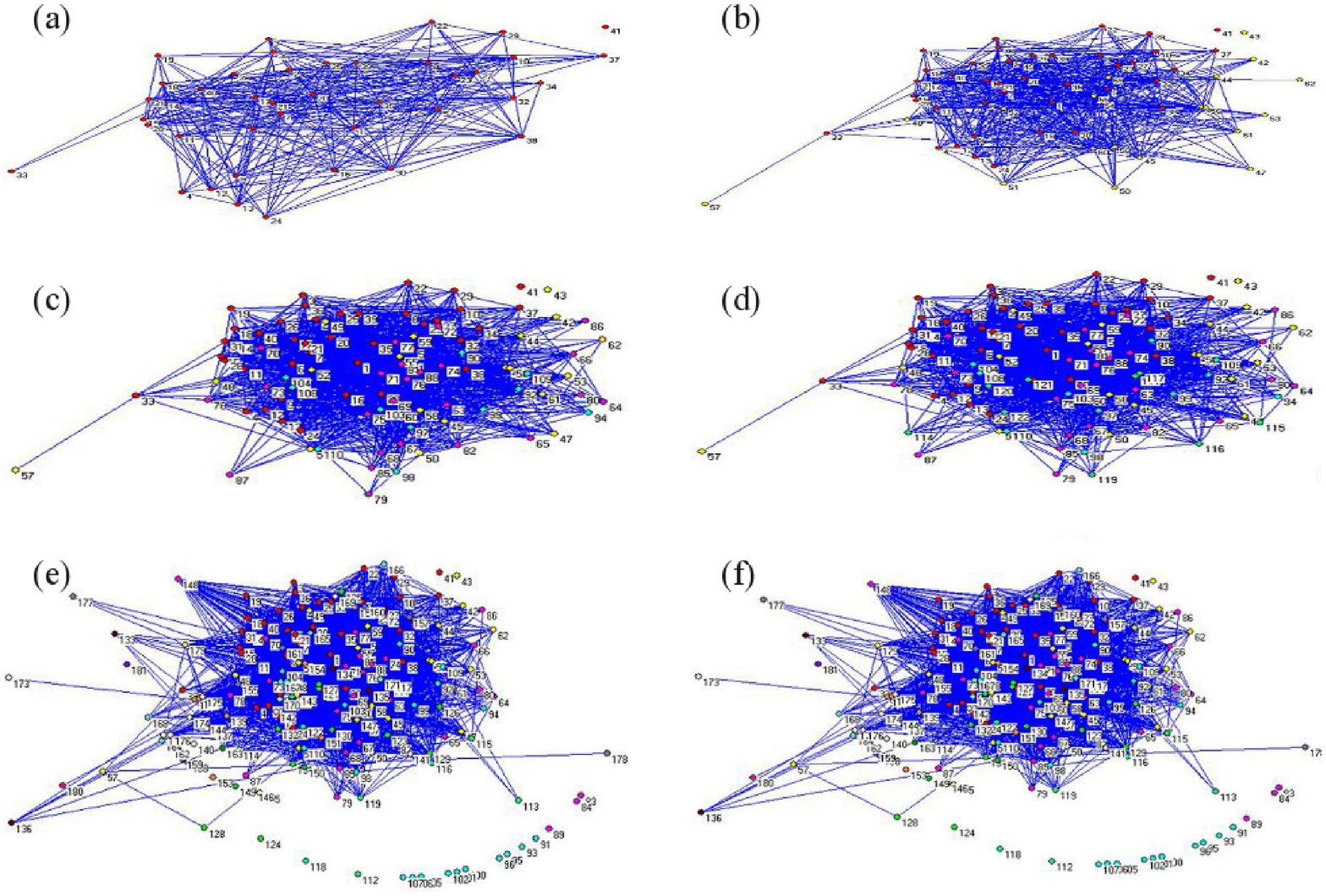,width=6.5in}
	\caption{The evolution graph for $\info_1$ during: (a)~Hour 1; (b)~Hour 2; (c)~Hour 4; (d)~Hour 5; (e)~Hour 29; and (f)~Hour 30.}\label{fig:evol1}
\end{figure*}

\begin{figure*}[htb]
\centering\epsfig{file=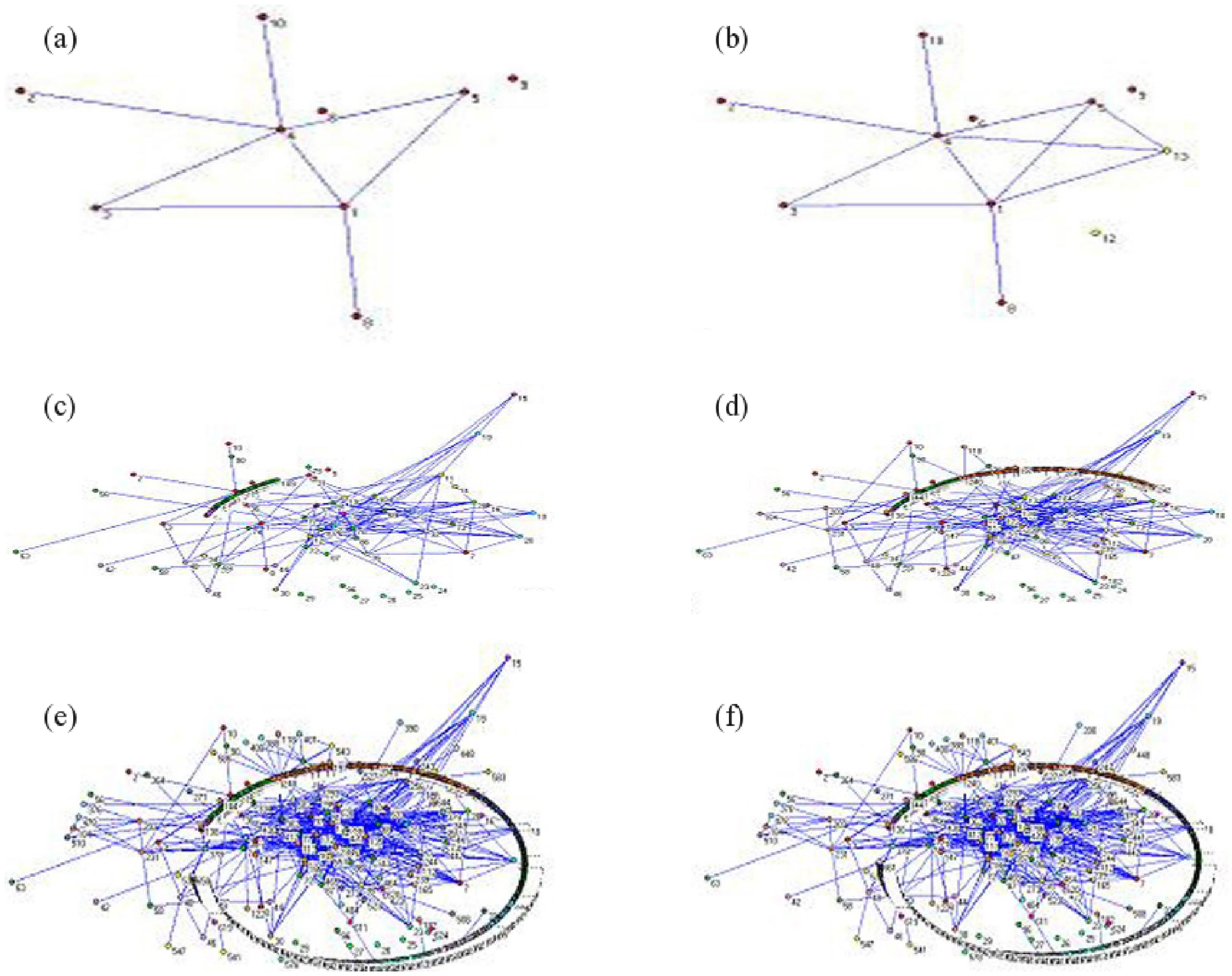,width=6in}
	\caption{The evolution graph for $\info_5$ during: (a)~Hour 1; (b)~Hour 2; (c)~Hour 15; (d)~Hour 16; (e)~Hour 29; and (f)~Hour 30.}\label{fig:evol2}
\end{figure*}

\begin{figure*}[htb]
\centering\epsfig{file=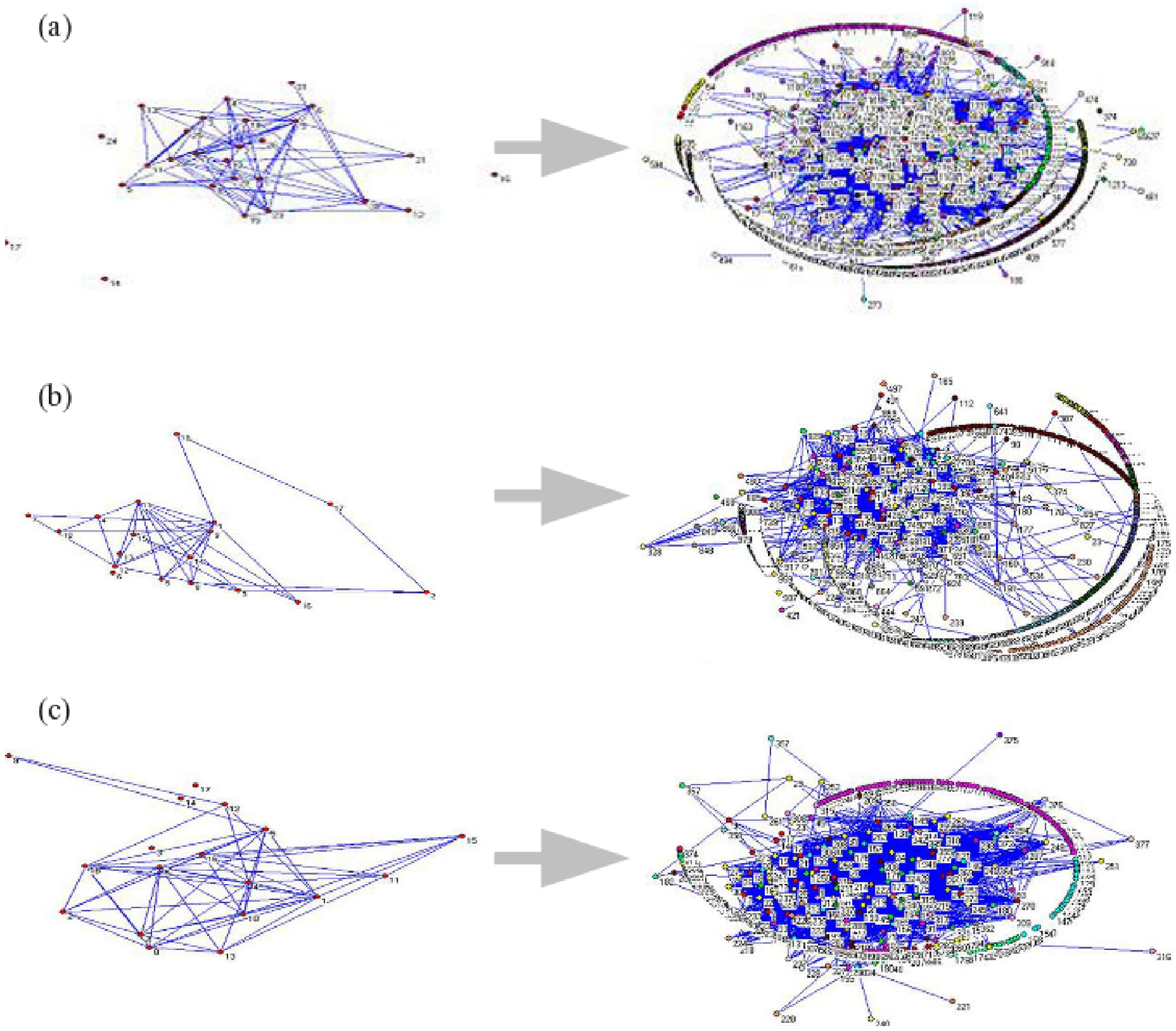,width=6in}
\caption{The respective evolution graphs of (a)~$\info_3$, (b)~$\info_4$, and (c)~$\info_6$ showing only Hour~1 and Hour~30.}\label{fig:appendix}
\end{figure*}

Figures~\ref{fig:evol1} and~\ref{fig:evol2} show the respective evolution graphs of~$\info_1$ and $\info_5$ during the three phases. During the expansion phase (Figures~\ref{fig:evol1}a--b and~\ref{fig:evol2}a--b), $\info_1$ is visible from the upcoming queue. Most of the first few diggers are direct friends of the submitter~$v_0$, thus the visualization shows a connected graph. During the front-page phase (Figure~\ref{fig:evol1}c--d and~\ref{fig:evol2}c--d), $\info_1$ became more visible to the other member of~$\SN$ causing it to accumulate votes at a faster rate. Some of the diggers are disconnected from the largest connected component. During the saturation phase (Figure~\ref{fig:evol1}e--f and~\ref{fig:evol2}e--f), $\info_1$ ages. This is characterized by the slow down of the accumulation of votes. The respective evolution graphs for $\info_3$, $\info_4$, and $\info_6$ are shown in Figure~\ref{fig:appendix}.

\section{Conclusion and Recommendations}

We looked at the effect of promotion in the speed and width of spread of information on the Internet by tracking the diffusion of news articles over a social network. We defined the speed~$\speed$ of spread as the number of readers that the news has reached in a given time, while width of spread means how far the story has travelled from the news originator within the social network. After analyzing six stories in a 30-hour time span, we found out that the lifetime of a story's popularity among the members of the social network has three phases, namely, the expansion, the front-page, and the saturation phases. The expansion phase starts when a story is published and the article spreads from a source node to nodes within a connected component of the social network. The front-page phase happens when a news aggregator promotes the story in its front page resulting to the story's faster rate of spread among the connected nodes, while at the same time spreading the article to nodes outside the original connected component of the social network. Saturation phase happens when the story ages and its rate of spread within the social network slows down suggesting popularity saturation among the social members. Within these three phases, we observed minimal changes on the width of information spread as suggested by relatively low increase of the width of the spread's diameter within the social network. The contribution of this paper is that this provides various stakeholders, such as the scientists and engineers, a first-hand empirical data to help them in modeling, designing, and improving the current web-based services. Similarly, the IT trainors and educators may find the information here useful for designing and improving curricula, as well as in improving the web-enabled deployment of knowledge and online evaluation of skills.

\section{Acknowledgments}

This research effort is funded by the Institute of Computer Science, \UPLB. 
\bibliography{journal}
\bibliographystyle{plainnat}


\end{document}